\documentstyle[aps,pre]{revtex}
\begin{document}
\draft


\title{Resonant Raman Scattering by
Charge Density and Single Particle Excitations in Semiconductor Nanostructures:
A Generalized Interband-Resonant Random-Phase-Approximation Theory}

\author{Daw-Wei Wang and S. Das Sarma} 

\address{Department of Physics, University of Maryland, College Park, Maryland 20742-4111}

\date{\today}
\maketitle
\pagenumbering{arabic}

\vspace*{1.5cm}

\begin{abstract}

We develop a generic theory for the
resonant inelastic light (Raman) scattering by a conduction band quantum 
plasma taking into account the presence of the filled 
valence band in doped 
semiconductor nanostructures within a generalized resonant
random phase approximation (RPA). Our generalized RPA theory explicitly 
incorporates the two-step resonance process where an electron from 
the filled valence band is first excited by the incident photon into 
the conduction band before an electron from the conduction band falls 
back into the valence band emitting the scattered photon. We show that 
when the incident photon energy is close to a resonance 
energy, i.e. the valence-to-conduction band gap of 
the semiconductor structure, 
the Raman scattering spectral weight at single particle excitation
energies may be
substantially enhanced even for long wavelength excitations, and 
may become comparable to the spectral weight of collective charge
density excitations (plasmon). Away from resonance, i.e.
when the incident photon energy is different from the band gap energy, 
plasmons dominate the Raman scattering spectrum.
We find no qualitative difference in the resonance effects on the
Raman scattering spectra among systems of different dimensionalities
(one, two and three) within RPA.
This is explained by the decoherence effect of the resonant
interband transition on the collective motion of conduction band electrons.
Our theoretical calculations agree well (qualitatively and 
semi-quantitatively) with the available experimental results,
in contrast to the standard nonresonant RPA
theory which predicts vanishing long wavelength Raman spectral weight 
for single particle excitations.
\end{abstract}


\vfill\eject
\vskip 1pc
\newcommand{\bfk}{\mathbf{k}\it}
\newcommand{\bfp}{\mathbf{p}\it}
\newcommand{\bfq}{\mathbf{q}\it}
\newcommand{\bfr}{\mathbf{r}\it}
\newcommand{\bfA}{\mathbf{A}\it}
\newcommand{\bfx}{\mathbf{x}\it}
\section{Introduction}
In recent years the elementary electronic excitation spectra of 
a variety of doped semiconductor nanostructures,
such as two dimensional (2D) quantum wells (QW) heterostructures, 
superlattices, and 
more recently, one dimensional (1D) quantum
wire (QWR) systems, have been studied extensively both experimentally 
\cite{6,rrs_exp96,Jusserand00,1,2,3,4,5,7,8,10}
and theoretically 
\cite{Dassarma99,Sassetti98,Wang00,wang01,11,12,13,14}. Rich experimental
spectra of the elementary electronic 
excitations (such as charge density 
excitations (CDE), spin density excitations (SDE), and 
single particle excitations (SPE), for both 
intrasubband and intersubband excitations) in these systems 
are typically experimentally investigated 
by using the resonant Raman scattering (RRS) technique, which is
a powerful and versatile spectroscopic tool to study
interacting electron systems.
In the RRS experiment, 
external photons are absorbed at 
one frequency and momentum, $\omega_i$ and $\bfk_i$, and emitted
at another, $\omega_f$ and $\bfk_f$, creating one or more particle-hole 
pair (or collective) excitations in the 
conduction band. The energy and momentum difference between
the incident photon and the scattered photon is Stokes shift, indicating
the dispersion of the relevant elementary electronic excitation created
in the system.
In the so-called polarized RRS geometry with the incident 
and scattered photons having the same polarization, the excited electrons 
have no spin flips during the scattering process, which therefore corresponds to the
elementary charge density excitations of the system. At low temperatures
(which is of interest to us in this paper) there is no real absorption 
of elementary excitations by the incident photon and the anti-Stokes 
line is not of any importance. (We use $\hbar=1$ throughout this paper.)

In the standard theory \cite{11,12,13,14,15}, 
which ignores the role of the valence
band and simplistically assumes the external photons 
to be interacting entirely with 
conduction band electrons, the polarized RRS intensity 
is proportional to the dynamical structure factor \cite{15,fetter}
of the conduction band electrons and 
therefore has strong spectral peaks at the 
collective mode frequencies at the wavevectors defined by the 
experimental geometry.
The dynamical structure factor peaks correspond to the 
poles of the reducible density response function, which are given by
the collective CDEs (plasmons) of the electron system 
in the long wavelength limit.
In particular, the single 
particle electron-hole excitations, which are at the poles of the 
corresponding irreducible response function, carry no long wavelength spectral 
weight (about three orders of magnitude weaker than the CDE spectral
weight at the typical wavevector, $10^5$ cm$^{-1}$, accessible in RRS 
experiments) in the density 
response function (according to the $f$-sum rule \cite{15}). 
The SPE therefore should
\textit{not}, as a matter of principle, show up in the polarized RRS spectra
in any dimensions. 
The remarkable experimental fact is that
there is always a relatively weak (but quite distinct) SPE 
peak in the observed polarized RRS spectra 
in addition to the expected 
CDE peak. This experimental presence of SPE peak in RRS cannot 
be explained by the standard theory, which, however,
does give the correct mode dispersion energy for both the
CDE and the SPE, but fails to explain why the SPE spectral weight is strongly
enhanced in the RRS experiments. This
puzzling feature \cite{6,rrs_exp96,14} of 
an ubiquitous anomalous SPE peak in addition to 
the expected CDE peak (or equivalently a two-peak structure) 
occurs in one, two, 
and even in three dimensional doped semiconductor nanostructures \cite{1}.
It exists in low dimensional semiconductor systems both for intrasubband and 
intersubband excitations. 

Many theoretical proposals \cite{Dassarma99,Sassetti98,Wang00,wang01,16,22}
have been made to explain this two-peak RRS puzzle.
$Ad$ $hoc$ proposals \cite{2,16} have been made
in the literature that perhaps a serious breakdown of momentum or wavevector
conservation (arising, for example, from scattering by random impurities) is
responsible for somehow transferring spectral weight from large to small
wavevectors, because the usual linear response theory predicts
that at very large wavevectors (an order of magnitude larger than the 
experimentally used RRS wavevectors), 
where the CDE mode is severely Landau damped, the dynamical 
structure factor should contain finite SPE spectral weight corresponding to 
high energy electron-hole excitations. 
Apart from being completely $ad$ $hoc$,
this proposal also suffers from any
lack of empirical evidence in its support --- in particular, the observed 
anomalous SPE peak in the RRS spectra does not correlate with the 
strength of the impurity scattering in the system.
We have recently systematically analyzed~\cite{wang01} all the proposed
mechanisms within the non-resonant RRS theory (i.e. without incorporating 
the valence band in the theory, assuming simply the inelastic 
light scattering process to be entirely confined to the conduction 
band free carrier system) leading to the conclusion that none of 
the proposed nonresonant mechanisms can explain the ubiquitous 
two-peak (the lower energy SPE peak and the higher energy CDE peak) 
structure of the observed RRS spectra.

We have recently reported \cite{Dassarma99} in a short letter a new
\textit{resonant} RRS theory by generalizing the nonresonant
RPA theory to include the filled valence band in the semiconductor,
reflecting the two-step resonant nature of the RRS process. The purpose
of the current article is to provide the details of our resonant RRS 
theory, and more importantly, to present RRS results for 2D and 3D
systems which have not been discussed earlier in the literature
(our earlier letter \cite{Dassarma99} presented only 1D RRS results).
The observed experimental RRS phenomenology in 1D, 2D, and 3D systems 
being very similar qualitatively, our generic interband-resonant
RRS theory, as reported herein, provides the conceptual theoretical
foundation for understanding RRS spectroscopy in doped semiconductor
structures.

In this context, we emphasize that the striking similarity of the
experimental RRS spectra in one, two, and three dimensional semiconductor 
systems suggests that the problem (namely, the two peak nature of 
the RRS spectra with the conspicuous presence of the "forbidden" 
SPE peak) is not specific to 1D systems, where
our earlier theory \cite{Dassarma99} was applied.
The ubiquitousness of the strong SPE spectral weight
in the RRS experiment (independent of system dimensionality,
dependent only on the resonant nature of the experiment)
suggests that the theoretical explanation for this puzzle must 
arise from some generic physics underlying 
RRS itself, and cannot be explained by the non-generic and 
manifestly system-specific theories which have been made occasionally 
in the literature. 
The resonant RRS theory presented herein (and in our short letter)
provides a \textit{generic} explanation for the two-peak structure
of the RRS spectra by establishing that the so-called low energy
anomalous SPE feature in the RRS spectrum arises entirely from the resonant
two-step nature of the RRS experiment, and cannot be the explained within
any non-resonant theory.

In this paper we provide (within the resonant RPA scheme) the
\textit{compellingly generic} theory for RRS
experiments by including the valence band electrons during the scattering 
processes for one, two and three dimensional semiconductor 
systems, following our earlier short paper \cite{Dassarma99} on 1D systems. 
We find that the RRS spectral weight at SPE energy is a strong 
function of the resonance condition ---
the SPE spectral weight is substantially enhanced when the incident 
photon frequency is near the semiconductor band gap
resonance energy, and decreases drastically
away from the resonance. It is
important to emphasize that this feature of our theory agrees with 
experimental observations --- the anomalous SPE peak exists only 
around resonance
and its spectral strength decreases off resonance.
Our results show similar qualitative behavior for the RRS 
spectra in one, two and three dimensional systems.

One dimensional systems \cite{hubbard} actually pose a special 
(and subtle) problem with respect to understanding the two-peak RRS spectra
because 1D electron systems are generically \cite{schulz,ll} Luttinger
liquids (i.e. non-Fermi liquids) which have no quasiparticle
(SPE) excitations whatsoever. The elementary excitations in 1D electron systems are bosonic spinon and holon collective modes. It is therefore
conceptually problematic to comprehend how an anomalous "SPE" feature
can arise in 1D semiconductor quantum wire RRS spectra as has been 
experimentally observed \cite{6,rrs_exp96,Jusserand00,8,10}. 
The issue of understanding 1D RRS spectra from a Luttinger liquid 
viewpoint has been recently discussed \cite{Sassetti98,Wang00,wang01} 
in the literature, and we refrain from further discussing this point 
in this article since this is beyond the scope of our work. 
In particular, our use of generalized RPA enables us to develop 
a unified consistent theory of resonant RRS in arbitrary dimensions 
(including 1D), and the Luttinger liquid nature of 1D quantum 
wires is not of any relevance in our theory. We mention, however, 
that a complete Luttinger liquid theory of 1D RRS experiments has 
recently been developed \cite{Wang00}, and this Luttinger liquid 
theory builds on the resonant nature of our work presented in this article. 

The rest of this paper is organized as follows: in Sec. II we describe the 
theory of nonresonant and resonant Raman scattering process in RPA; in
Sec. III we present and discuss our calculated RRS results 
for one, two, and three dimensional
GaAs semiconductor systems; 
we then summarize our work in Sec. IV. All the results shown in this 
paper are for GaAs-based systems, but obviously the theory 
applies to any direct band gap semiconductor material.
\section{Theory}
In Fig. 1(a) we depict the schematic diagram \cite{Dassarma99,17,18,19} for the
two step process (steps 1 and 2 in the figure) involved in the
polarized resonant Raman scattering spectroscopy at the
$E_0+\Delta_0$ direct gap of electron doped GaAs system
\cite{19} where an electron in the valence
band is excited by the incident photon into an excited (i.e. above the Fermi
level) conduction band state, leaving a valence band hole behind (step 1);
then an electron from inside the conduction band Fermi surface recombines with
the hole in the valence band (step 2), emitting an outgoing photon with 
an energy and momentum (Stokes) shift. 
The RRS process is a two-step process involving steps 1 and 2 
with the net result of there being an elementary electron 
excitation created in the conduction band through 
intermediate valence band states as shown in Fig. 1. The 
non-resonant approximation to RRS ignores the intermediate 
valence band states and approximates the RRS process to be 
taking place entirely within the conduction band of the 
system, as shown by the step 3 in Fig. 1. The whole point 
of the theory \cite{Dassarma99} developed in this paper 
is that the non-resonant step 3 is not equivalent to the 
resonant scattering involving steps 1 and 2. Note that the 
resonant process involving steps 1 and 2 explicitly 
depends on the incident photon energy, while
the non-resonant approximation depicted in step 3 depends 
only on the energy difference between the incident and the
scattered photons and not on the incident photon energy. This difference
turns out to be crucial in the RRS theory, and the resonance 
condition in the incident photon energy gives rise to 
the anomalous SPE-like feature in the RRS spectra as shown below.
Electron spin is conserved throughout the scattering processes since
we are considering only the polarized geometry where no spin flip occurs.
As mentioned before we use the random phase approximation (RPA) in our 
theory taking care to generalize it to the resonant situation 
involving steps 1 and 2. In the RPA one neglects all 
exchange-correlation effects (e.g. self-energy and vertex 
corrections due to electron-electron interaction), including only 
the long range Coulomb interaction $V_c(\bfq)$ 
in the dynamical screening by the electron system so as 
to correct the noninteracting irreducible response function 
to the reducible response function. Following a preliminary 
discussion of the Coulomb interaction in 1, 2 and 3 
dimensional semiconductor system in subsection A below, 
we then develop the nonresonant and the resonant RRS 
theories in sections B and C respectively. Our theory 
is entirely within the effective mass approximation, and 
we parameterize the electron system in the semiconductor 
by electron ($m_e$) and hole ($m_h$) effective masses 
corresponding to the top (bottom) of the conduction 
(valence) band and by a background lattice dielectric 
constant $\varepsilon_0$.
\subsection{Coulomb Interaction}
The realistic (bare) Coulomb interaction in the 
artificially confined semiconductor
nanostructures depends strongly
on the confinement geometry of the systems. 
In the bulk 3D semiconductor materials, the 
unscreened Coulomb interaction has a long-ranged $1/r$ decay 
in the real space, and has
the following Fourier transform in momentum space:
\begin{equation}
V^{\rm{3D}}_c(\bfq)=\frac{e^{\rm 2}}{\varepsilon_0}
\int \frac{{d}\bfr^{\rm{3}}}{|\bfr|}e^{{i}\bfq\cdot\bfr}
=\frac{\rm{4}\pi\it e^{\rm 2}}{\varepsilon_{\rm{0}} |\bfq|^{\rm{2}}}, 
\end{equation}
where $e$ is the electron charge and $\varepsilon_0$ is the 
dielectric constant of the background material (about 12 in the
GaAs semiconductor system). We use the static ($\varepsilon_0$) lattice dielectric
constant in our theory rather than the more conventional high frequency 
($\varepsilon_\infty$) dielectric constant in defining the 
Coulomb interaction, $V_c(\bfq)$, because inclusion of 
$\varepsilon_0$ is known to approximately account for the 
polaronic electron-phonon interaction in the system, which 
is rather weak in GaAs because of its low Fr$\ddot{\rm o}$hlich 
coupling constant ($\sim 0.07$).
In the 2D semiconductor quantum well system, modern fabrication techniques
have produced very narrow 2D wells (of nanostructure size $<100$
\AA\ in GaAs in the confinement direction), 
leading to an almost pure 2D electron system.
It is therefore a good approximation to assume the well width
to be zero in our calculation, giving a 2D Fourier transform 
of the Coulomb interaction:
\begin{equation}
V^{\rm{2D}}_c(\bfq)=\frac{e^{\rm 2}}{\varepsilon_0}
\int \frac{{d}\bfr^{\rm{2}}}{|\bfr|}e^{{i}\bfq\cdot\bfr}
=\frac{\rm{2}\pi \it{e}^{\rm 2}}{\varepsilon_{\rm{0}} |\bfq|}.
\end{equation}
Inclusion of the confinement wavefunction effect in the theory is 
straightforward and leads to a form factor $f(\bfq)$ ($<1$) multiplying
$V_c^{\rm{2D}}(\bfq)$ in the theory.
For the 1D semiconductor quantum wire system, we have to consider the realistic
finite width of the wire (i.e. the relevant 1D form factor effect) 
because the 1D Fourier transform of $1/r$ potential
(i.e. $\int dr e^{iqr}/|r|$) diverges logarithmically requiring 
regularization by a length cut-off associated 
with the typical confinement size. 
Therefore the Coulomb interaction
for the finite width quantum wire is obtained by taking the
expectation value of the 2D Coulomb interaction (assuming the width
in the $z$ direction to be zero for simplicity as in our 2D model in Eq. 2) 
over the confinement
wavefunction along the transverse direction ($y$) of
the wire. We then have~\cite{benhu} the following
Coulomb interaction matrix element in the 1D QWR structure of finite width:
\begin{eqnarray}
V^{\rm{1D}}_{c,ij}(q)&=&\frac{e^{\rm 2}}
{\varepsilon_0}\int_{-\infty}^{\infty}\it{}dy\,dy'
\int_{-\infty}^{\infty}dx 
\frac{e^{-iqx}|\phi_i(y)|^{\rm 2\it}|\phi_j(y')|^{\rm 2\it}}
{\sqrt{x^{\rm 2\it}+(y-y')^{\rm 2\it}}}
\nonumber\\
&=&\frac{2e^{\rm 2}}{\varepsilon_0}\int_{-\infty}^{\infty}\it{}dy\,dy'
|\phi_i(y)|^{\rm 2\it}|\phi_j(y')|^{\rm 2\it}
K_{\rm 0}(q|y-y'|^2),
\end{eqnarray}
for interaction between electrons of subband $i$ and subband $j$.
$\phi_i(y)$ is the electron wavefunction of $i$th subband of the
QWR along the transverse direction. In this paper
we assume that only the lowest ($i=1$) ground conduction subband is occupied (i.e.
subband spacing $\Delta E_{12}> E_F$ at zero temperature and all the higher 
energy subbands are empty) and neglect any intersubband transition, 
so that the subband index $i=1$ throughout and 
will not be explicitly shown. 
$K_0(x)$ in Eq. (3) is the zeroth-order
modified Bessel function of the second kind.
The exact form of wavefunction $\phi(y)$
depends on the confinement geometry of the QWR system.
For simplicity we assume the QWR confinement potential to be the 1D infinite
square well in the $y$ direction. This turns out to be a 
good approximation for the electrostatic gate-controlled confinement
in the presence of 
the self-consistent Hartree potential due to the free electrons themselves
\cite{benhu}. The the confinement wavefunction $\phi(y)$ is
(for the ground subband):
\begin{eqnarray}
\phi(y)=\left\{ \begin{array}{ll}
       \sqrt{\frac{\displaystyle 2}{\displaystyle a}}
          \cos\left(\frac{\displaystyle \pi y}{\displaystyle a}\right)\ 
          & \mbox{if} -a/2<y<a/2 \\
       0 & \rm{otherwise},
       \end{array}\right.
\end{eqnarray} 
where $a$ is the wire width in the $y$ direction. Using Eqs. (3) and (4)
we can numerically calculate the effective 1D Coulomb interaction 
\cite{benhu} for
the semiconductor QWR system. Unlike the power-law behavior of Coulomb
interaction in the higher dimensions (Eqs. (1) and (2)),
$V^{\rm{1D}}_{c}(q)$ has a weak logarithmic divergence, 
$-2{e}^2\ln(qa)/\varepsilon_0$, in the long wavelength limit 
($q\rightarrow 0$). Because of this logarithmic dependence of $V_c(q)$
on $q$ (as $q\rightarrow 0$), the precise value of the wire width 
($a$) is not particularly important in our theory, making our simple 
infinite square well approximation a reasonable one for our purpose.
\subsection{Nonresonant Raman scattering}
In the presence of an external photon field 
the interacting Hamiltonian between
the free electron gas and the radiation field is assumed to be obtainable
from the standard gauge-invariant prescription~\cite{sakurai,23}, 
$\bfp\rightarrow\bfp-e\bfA/c$, where
$\bfA$ is the radiation field (photon) vector 
potential operator and $c$ the speed of light.
The Hamiltonian including the radiation field and the
electrons (i.e. the free carriers induced by doping) 
in the semiconductor conduction band can therefore be
written as (we neglect the spin-photon interaction considering only
polarized RRS spectra where spins do not play any explicit role)
\begin{eqnarray}
H&=&\sum_i^N\frac{1}{2m_e}\left(\bfp_i-e\bfA(\bfx_i,t)/c\right)^2+
\sum^N_{ij,i>j}
V_c(\bfx_i-\bfx_j)
\nonumber\\
&=&\underbrace{\sum_i^N\left[\frac{\bfp_i^{\rm 2}}{2m_e}
+\sum^N_{j<i}V_c(\bfx_i-\bfx_j)\right]}_{\displaystyle H_0}
+\underbrace{\sum_i^N\left[-\frac{e}{m_ec}\bfp_i\cdot\bfA(\bfx_i,t)
+\frac{e^{\rm 2}}{\rm 2\it m_ec^{\rm 2}}
\bfA(\bfx_i,t)^{\rm 2}\right]}_{\displaystyle H_I}
\end{eqnarray}
in the effective mass approximation with $m_e$ being the 
effective electron mass of the semiconductor conduction band. We have made
the transverse gauge choice \cite{sakurai}, 
$\nabla\cdot\bfA(\bfx_i,t)=\rm{0}$, for the radiation
field, leading to $\bfp_i\cdot\bfA=\bfA\cdot\bfp_i$ as used in Eq. (5). 
$H_0$ is the Hamiltonian of electrons
interacting with Coulomb potential without the radiation field,
and $H_I$ is the electron-photon 
interaction Hamiltonian which plays a crucial role in the Raman 
scattering problem. Figs. 1(b) and 1(c) correspond to
the scattering processes induced by
the linear ($\bfp\cdot\bfA$) term and the quadratic ($\bfA^{\rm 2}$) 
term respectively in the second quantization representation. 
The $\bfp\cdot\bfA$ term 
creates and annihilates one photon in the state it acts on, having 
no contribution to the scattering rate in the first order time-dependent
perturbation theory since there is
no net change of photon numbers.
The quadratic $\bfA^{\rm 2}$ term, on the other hand,
gives a non-vanishing first order contribution to the 
scattering rate because photons are created and annihilated at the same time
in such scattering processes as shown in Fig. 1(c).
In principle the second order 
contribution of $\bfp\cdot\bfA$ term in the time-dependent perturbation theory
is of the the same order as the first order contribution from the 
$\bfA^{\rm 2}$ term as a simple power counting in the coupling
constant $e/c$ shows. This second order contribution,
which plays a role in the RRS phenomenon, will be studied and
discussed in more details in the next section.
We can simply neglect this ($\bfp\cdot\bfA$) term in $H_I$ if we are
interested only in the $non$resonant Raman scattering regime, 
either because the incident
photon energy $\omega_i$ is off-resonance i.e. far from
the direct band gap, $E_g^0$ ($\sim$
1.5 eV in GaAs), or because we only want to consider a nonresonant 
process as in step 3 in Fig. 1(a). The $\bfA^{\rm 2}$ term, 
being a scalar field operator which commutes with the electron field,
$\psi(\bfx)$, leading to the
perturbative Hamiltonian, $H_I$ (neglecting the $\bfp\cdot\bfA$
term) being
proportional to the electron density operator, 
$n(\bfx)=\sum_s\psi_s^\dagger(\bfx)\psi_s(\bfx)$.
The nonresonant (corresponding to the step 3 process in Fig. 1) 
Raman scattering intensity at frequency shift $\omega$ and momentum 
transfer $\bfq$ therefore can be calculated
from the dynamical structure factor (the imaginary part of the density
response function) in the linear response theory~\cite{15,fetter}:
\begin{equation}
\frac{d^2\sigma}{d\Omega d\omega}\propto -\mathrm{Im}
\Pi(\bfq,\omega)=\rm{Im}\left[\it i\int_{\rm{0}}^\infty dt e^{i\omega t}
\langle [n^\dagger(\bfq,t),n(\bfq,\rm{0})]\rangle_{\rm{0}}\right],
\end{equation}
where $\langle\cdot\cdot\cdot\rangle_0$ is the ground 
state expectation value, and 
$n(\bfq,t)$ is the electron density operator, 
$n(\bfq,t)=\sum_{\bfk,s}c^\dagger_{\bfk+\bfq,s}(t)c_{\bfk,s}(t)$, with 
$c^\dagger_{\bfk,s}(c_{\bfk,s})$ the electron creation(annihilation)
operator for momentum $\bfk$ and spin $s$.
In the standard many-body theory, this (reducible) response function
can be obtained by the reducible set of polarization diagrams \cite{15,fetter}
(Dyson's equation, see Fig. 2) formed 
by the irreducible conduction band polarizability, $\Pi_{0}(\bfq,\omega)$,
for the scattering process
where one has an electron and a hole in the conduction band:
\begin{eqnarray}
\Pi(\bfq,\omega)&=&
\Pi_{0}(\bfq,\omega)+\rm{\Pi_{0}}(\bfq,\omega)V_c(\bfq)
\rm{\Pi_{0}}(\bfq,\omega)+\cdot\cdot\cdot
\nonumber\\
&=&\frac{\Pi_{0}(\bfq,\omega)}{1-V_c(\bfq)\rm{\Pi_{0}}(\bfq,\omega)}
=\frac{\Pi_{0}(\bfq,\omega)}{\epsilon(\bfq,\omega)},
\end{eqnarray}
where $\epsilon(\bfq,\omega)\equiv \rm{1}\it 
-V_c(\bfq)\rm{\Pi_{0}}(\bfq,\omega)$ 
is the dynamical dielectric function.

In the random phase approximation (used in this paper),
the irreducible polarizability, $\Pi_{0}(\bfq,\omega)$, 
is approximated by the noninteracting electron-hole bubble,
$\Pi_{0}^{\mathrm{RPA}}(\bfq,\omega)$, without 
any self-energy or vertex correction. 
RPA is known to be a good approximation
\cite{11,12,13,17,18,19} 
in two- and three-dimensional electron systems for calculating plasmon
(or CDE) properties. It 
is also a good approximation for collective mode dispersion
in one-dimensional electron systems and gives a 1D plasmon dispersion 
which agree with 
the exact Luttinger liquid theory \cite{benhu}.
The expression of
$\Pi_{0}^{\mathrm{RPA}}(\bfq,\omega)$ for a $d$-dimensional system is
\begin{eqnarray}
\Pi^{\mathrm{RPA}}_{0}(\bfq,\omega)
&=&\frac{-2i}{(2\pi)^{d+1}}\int d\nu d\bfp\; G_{\rm 0}(\bfp,\nu)
G_{\rm 0}(\bfp+\bfq,\nu+\omega)
\nonumber\\
&=&\frac{-2}{(2\pi)^d}\int d\bfp\
\frac{n_{\rm 0}(\bfp)-n_{\rm 0}(\bfp-\bfq)}
{\omega+i\gamma-\bfp^{\rm 2}/\rm{2}\it{m_e}+
(\bfp-\bfq)^{\rm 2}/\rm{2}\it{m_e}},
\end{eqnarray}
where $G_0(\bfp,\nu)$ is the bare conduction 
band electron Green's function and 
$n_0(\bfp)=\theta(k_F-|\bfp|)$ is the zero temperature
noninteracting momentum distribution function of conduction band electrons.
$\gamma$ is a phenomenological damping term associated with 
impurity scattering (or other broadening mechanism), which is
taken to be small ($\gamma\ll E_F$) in our numerical calculation. 
The damping term, $\gamma$, introduces finite widths to the 
spectral peaks in the dynamical structure factor of Eq. 
(6), but does not affect the peak position and spectral 
weight in any significant method.
The imaginary part of the irreducible polarizability 
$\Pi_{0}(\bfq,\omega)$ (which is now approximated by 
$\Pi_{0}^{\mathrm{RPA}}(\bfq,\omega)$ in our paper) gives rise to the single
particle excitation, which is typically very small at long wavelengths
due to the dynamical screening effect of Eq. (7). 
In Fig. 3 we show as shaded regions the SPE continua 
(where Im$\Pi_{0}(\bfq,\omega)\neq\rm 0$)
within RPA for one, two, and three dimensional 
systems. 
Note that, in contrast to 2D and 3D systems, the 1D SPE continuum 
is very restricted in the long wavelength limit ($q\ll k_F$).
In higher dimensions, the SPE continuum is
gapless for any finite wavevector smaller than $2k_F$, but it is gapped in
1D due to energy-momentum conservation induced phase space restriction.
Using Eqs. (6)-(8) we can calculate the nonresonant Raman scattering spectra
and the plasmon (CDE) dispersion (shown in Fig. 3)
to compare with the experimental results and the resonant theory
results discussed below. The calculated spectra are
shown in Figs. 4(a)-(c) for one, two and three dimensional systems
respectively. We discuss these results in details in Sec. III.
\subsection{Resonant Raman scattering}
We now consider the full resonance situation (step 1 and 2 in Fig. 1)
including the
valence band which obviously 
\cite{Jusserand00,Dassarma99,Sassetti98,Wang00,23}
plays a crucial role in the RRS
experiment because the external photon energy must be approximately equal the
$E_0+\Delta_0$ direct gap for the experiment to succeed.
In the RRS process the incident
photon is absorbed and a scattered photon with the appropriately shifted
frequency (and wavevector) 
is emitted. Electron spin is conserved throughout the scattering
process. As discussed above,
there are two steps (steps 1 and 2 in Fig. 1(a)) 
involved in the polarized RRS spectroscopy and both
of these two steps of inelastic scattering result from
the $\bfp\cdot\bfA$ term of $H_I$ in Eq. (5) (see Fig. 1(c)).
When the incident photon frequency is equal 
to the direct band gap energy, $E_0$,
the second order "resonant" perturbative contribution of
the $\bfp\cdot\bfA$ term becomes important and comparable
to the first order contribution of $\bfA^{\rm 2}$ term, leading to an electron
interband transition between the conduction band and the valence band.
The interaction Hamiltonian of the RRS theory with external photon 
momentum, $\bfk$, and frequency, $\omega$, can be expressed in the second
quantization representation as
\begin{eqnarray}
H_{I}^{\bfk,\omega}&=&e^{-i\omega t}\sum_{\bfp,\sigma }
[c_{\bfp,\sigma}^{\dagger}(t)v_{\bfp-\bfk,\sigma}(t)+
v_{\bfp,\sigma}^{\dagger}(t)c_{\bfp-\bfk,\sigma}(t)] \nonumber \\
&+&e^{i\omega t}\sum_{\bfp,\sigma}[c_{\bfp,\sigma}^{\dagger}(t)
v_{\bfp+\bfk,\sigma}(t)+
v_{\bfp,\sigma}^{\dagger}(t)c_{\bfp+\bfk,\sigma}(t)],
\end{eqnarray}
with $c_{\bfk,\sigma}$ and $v_{\bfk,\sigma}$ being the annihilation 
operators of conduction band and valence band
electrons respectively. The electron-photon coupling vertex,
$\frac{-e}{m_e cL}\mathbf{p}\cdot\mathbf{\epsilon}$ (where $\epsilon$ is
the light polarization), has been assumed to be
constant for simplicity.
Applying the time-dependent perturbation theory
to the ground state, $|0\rangle$, 
characterized by a conduction band Fermi sea and
no holes in the valence band (at zero temperature),
we have the following transition amplitude from ground state, $|0\rangle$,
to the $n$th 
excited state, $|n\rangle$:
\begin{equation}
c_n(T)=\int_{-T/2}^{T/2} dt_1\int_{-T/2}^{t_1}dt_2\langle n|
H_{I}^{\bfk_f,\omega_f}(t_1)H_{I}^{\bfk_i,\omega_i}
(t_2)|\rm 0\rangle,
\end{equation}
where we have changed the time 
integration range from the conventional $\{0,T\}$
to $\{-T/2,T/2\}$ for the convenience of changing variables later.
By substituting the explicit form of $H_{I}$ and choosing the specific
channel of backward scattering (the so-called back-scattering geometry),
$\bfk_i=-\bfk_f=\bfq/\rm{2}$ 
and $\omega_{i,f}=\Omega\pm\omega/2$, without any loss of generality, 
we obtain the transition rate (ignoring excitonic and self-energy
effects), $W$, to be
\begin{eqnarray}
W&=&\lim_{T\rightarrow\infty}\frac{1}{T}\left|\sum_n c_n(T)\right|^2
\nonumber\\
&=&\lim_{T\rightarrow\infty}\frac{1}{T}
\sum_{\stackrel{\bfp_{\rm 1}\cdot\cdot\cdot \bfp_{\rm 4}}
{\sigma_1\cdot\cdot\cdot\sigma_4}}
\int_{-T/2}^{T/2}dt_{1}\int_{-T/2}^{t_1}dt_{2}\int_{-T/2}^{T/2}
dt'_1\int_{-T/2}^{t'_1}dt'_2
e^{i\Omega(t'_2-t'_1+t_1-t_2)}e^{i\omega(t_2'+t_1'-t_1-t_2)/2} \nonumber\\
&&\times\langle v^{\dagger}_{\bfp_{\rm 1}-\bfq/\rm{2},\sigma_{\rm 1}}(t_2')
v_{\bfp_{\rm 2},\sigma_{\rm 2}}(t_1')
v^{\dagger}_{\bfp_{\rm 3},\sigma_{\rm 3}}(t_1)
v_{\bfp_{\rm 4}-\bfq/\rm{2},\sigma_{\rm 4}}
(t_2)\rangle_0 \nonumber\\
&&\times\langle c_{\bfp_{\rm{1}},\sigma_{\rm 1}}(t_2')
c^{\dagger}_{\bfp_{\rm 2}-\bfq/\rm{2},\sigma_2}(t_1')
c_{\bfp_{\rm 3}-\bfq/\rm{2},\sigma_3}(t_1)
c^{\dagger}_{\bfp_{\rm 4},\sigma_{\rm 4}}(t_2)\rangle_0.
\end{eqnarray}
Since the valence band is completely 
filled in the ground state at zero temperature,
we have only one contraction of the valence band electron operators,
which is assumed to be noninteracting for simplicity: 
\begin{eqnarray}
&&
\langle v^{\dagger}_{\bfp_{\rm 1}-\bfq/\rm{2},\sigma_{\rm 1}}(t_2')
v_{\bfp_{\rm 2},\sigma_{\rm 2}}(t_1')
v^{\dagger}_{\bfp_{\rm 3},\sigma_{\rm 3}}(t_1)
v_{\bfp_{\rm 4}-\bfq/\rm{2},\sigma_{\rm 4}}(t_2)\rangle_0 \nonumber\\
&&=\langle v^{\dagger}_{\bfp_{\rm 1}-\bfq/\rm{2},\sigma_{\rm 1}}(t_2')
v_{\bfp_{\rm 2},\sigma_{\rm 2}}(t_1')\rangle_0
\langle v^{\dagger}_{\bfp_{\rm 3},\sigma_{\rm 3}}(t_1)
v_{\bfp_{\rm 4}-\bfq/{\rm 2},\sigma_{\rm 4}}(t_2)\rangle_0.
\nonumber\\
&&=\delta_{\bfp_{\rm 1},\bfp_{\rm 2}+\bfq/\rm{2}}\delta_{\sigma_1,\sigma_2}
\delta_{\bfp_{\rm 3},\bfp_{\rm 4}-\bfq/\rm{2}}\delta_{\sigma_3,\sigma_4}
e^{iE_v(\bfp_{\rm 1}-\bfq/\rm{2})(\it{t'}_{\rm 1}-\it{t'}_{\rm 2})}
e^{iE_v(\bfp_{\rm 3})(\it{t}_{\rm 1}-\it{t}_{\rm 2})},
\end{eqnarray}
where $E_v(\bfp)=-\bfp^{\rm 2}/\rm 2\it{m_v}$ is 
the kinetic energy of the valence band
electrons.
Setting new implicit time variables ($t_{1,2}\rightarrow\bar{t}_1\pm t_1/2$
and $t'_{1,2}\rightarrow\bar{t}_2\pm t_2/2$) and using the quasi-particle
approximation for the electron operator,
$c_{\bfp,\sigma}(\bar{t}-t/2)=c_{\bfp,\sigma}(\bar{t})
e^{-iE_c(\bfp)t/2}$, we can
obtain the transition rate, after evaluating the $t_1$ and $t_2$ integrals,
\begin{eqnarray}
W&=&\lim_{T\rightarrow\infty}\frac{1}{T}\int_{-T/2}^{T/2}d\bar{t}_1
\int_{-T/2}^{T/2}d\bar{t}_2
\,e^{i\omega(\bar{t}_2-\bar{t}_1)}
\nonumber \\
&&\times\sum_{\stackrel{\bfp_{\rm 1},\bfp_{\rm 2}}{\sigma_{\rm 1},\sigma_2}}
A^\ast(\bfp_{\rm 1},\bfq) A(\bfp_{\rm 2},\bfq)
\langle c_{\bfp_{\rm 1}+\bfq/\rm{2},\sigma_{\rm 1}}(\bar{t}_{\rm 2})
c^{\dagger}_{\bfp_{\rm 1}-\bfq/\rm{2},\sigma_{\rm 1}}(\bar{t}_{\rm 2})
c_{\bfp_{\rm 2}-\bfq/\rm{2},\sigma_{\rm 2}}(\bar{t}_{\rm 1})
c^{\dagger}_{\bfp_{\rm 2}+\bfq/\rm{2},\sigma_2}(\bar{t}_{\rm 1})\rangle_{\rm 0}
\nonumber \\
&=&\int_0^\infty dt\,e^{i\omega t}\langle 
N^\dagger(\bfq,t)N(\bfq,\rm{0})\rangle_{\rm 0},
\end{eqnarray}
where the resonant "density" operator, $N(\bfq,t)$, is defined to be
\begin{eqnarray}
N(\bfq,t)&=&-\sum_{\bfp,\sigma}A(\bfp,\bfq)c_{\bfp-\bfq/\rm{2},\sigma}(t)
c^{\dagger}_{\bfp+\bfq/\rm{2},\sigma}(t)
\nonumber\\
&=&\sum_{\bfp,\sigma}A(\bfp,\bfq)c^\dagger_{\bfp+\bfq/\rm{2},\sigma}(t)
c_{\bfp-\bfq/\rm{2},\sigma}(t)
\end{eqnarray}
for $\bfq\neq \rm{0}$ with the matrix element $A(\bfp,\bfq)$:
\begin{eqnarray}
A(\bfp,\bfq)&=&\frac{1}
{E_g-\Omega+(E_c(\bfp-\bfq/\rm{2})+\it E_c(\bfp+\bfq/\rm{2}))/2
-\it E_v(\bfp)+i\lambda}
\nonumber\\
&=&\frac{1/E_F}{E_{\omega}+(1+\xi)(\tilde{\bfp}^{\rm{2}}-\rm{1})
+\tilde{\bfq}^{\rm{2}}/\rm{4}+\it{i}\lambda/E_F}.
\end{eqnarray}
Here $E_{\omega}\equiv {E_F}^{-1}(E_g+(1+\xi)
E_{F}-\Omega)$ with 
$\xi\equiv m_c/m_v$; $\tilde{\bfp}\equiv \bfp/\mathit{k_F}$;
$\tilde{\bfq}\equiv \bfq/\mathit{k_F}$, and $E_F=E_c(k_F)$ is the Fermi
energy of the conduction band electrons.
$\lambda$ is a phenomenological broadening factor we introduce to include
roughly all possible broadening effects,
e.g. finite imaginary part of electron self-energy
(the quasi-particle life time), finite impurity or disorder scattering,
and any broadening or damping arising intrinsically from the photon 
field or the associated optical scattering.
We take $\lambda$ to be small (=0.02$E_F$)
in the numerical calculation.
Note that the phenomenological parameter, $\lambda$, is a resonance
broadening parameter (associated with the band to band process) to be contrasted
with the simple spectral broadening parameter, $\gamma$, of Eq. (8), which is 
purely a conduction band phenomenological parameter. In our leading order 
RRS theory $\lambda$ (of Eq. (15)) and $\gamma$ (of Eq. (8)) are 
completely independent phenomenological relaxation or damping terms 
(both of which should be small, $\gamma$ and $\lambda$ $\ll E_F$, 
for our leading order theory to be sensible). Calculation of $\gamma$ 
and $\lambda$ is beyond the scope of the leading order theory --- 
it is entirely possible that in a more complete theory including 
quasiparticle self-energy and vertex corrections as well as 
electron-impurity scattering and the electron-photon interaction, 
$\gamma$ and $\lambda$ will tern out to be related.

Comparing Eqs. (13)-(14) with Eq. (6), we 
find that the effect of resonance (i.e. photon induced interband transition)
on the 
conduction band electrons is the matrix element 
$A(\bfp,\bfq)$, which arises from the time difference between the
excitation of one electron from valence band to the conduction band
(step 1) and the 
recombination of another electron from inside the conduction band 
Fermi surface with the hole in the valence band (step 2). 
The resonance condition is parameterized by the dimensionless parameter, 
$E_\omega$, with $E_\omega=0$ being the precise resonance 
condition. In the following discussion we define "off resonance" 
as $|E_\omega|\gg 1$ and "near resonance" as $|E_\omega| \ll 1$. 
Off resonance 
the spectral weight decreases as $|E_\omega|^{-2}$ as can be seen from
Eq. (15).
Near resonance the singular properties of 
$A(\bfp,\bfq)$ enhances the spectral weight nontrivially.  
The calculation of the RRS spectrum is therefore reduced to the
evaluation of the correlation function of
Eq. (14), which in the resonant 
RPA approximation (i.e. neglecting
all vertex correction of the irreducible polarizabilities, see Fig. 5)
is obtained to be
\begin{equation}
W\approx -\mathrm{Im}\left[\Pi^{\rm RPA}_2(\bfq,\omega)+
\frac{\mathrm{\Pi}_1^{\rm RPA}(\bfq,\omega)
\mathrm{\bar{\Pi}_1^{\rm RPA}}(\bfq,\omega)
{\mathit{V_c}}(\bfq)}
{\epsilon(\bfq,\omega)} 
\right],
\end{equation}
where 
\begin{equation}
\Pi_2^{\rm RPA}(\bfq,\mathit{\omega})=\frac{-\mathrm{2}}{(\mathrm{2}
\mathit{\pi})^d}\int d\bfp\;
\frac{|A(\bfp,\bfq)|^{\rm{2}}(\it{n}_{\rm 0}(\bfp+\bfq/\rm{2})-
\it{n}_{\rm 0}(\bfp-\bfq/\rm{2}))}
{\omega +i\gamma -E_c({\bfp+\bfq/\rm{2}})+\it{E}_c({\bfp-\bfq/\rm{2}})}, 
\nonumber \\
\end{equation}
and
\begin{equation}
\Pi^{\rm RPA}_1(\bfq,\mathit{\omega})=\frac{-\mathrm{2}}{(\mathrm{2}
\mathit{\pi})^d}\int d\bfp\;
\frac{A(\bfp,\bfq)(\it{n}_{\rm 0}(\bfp+\bfq/\rm{2})-\it{n}_{\rm 0}
(\bfp-\bfq/\rm{2}))}
{\omega +i\gamma -E_c(\bfp+\bfq/\rm{2})+\it{E_c}(\bfp-\bfq/\rm{2})}, 
\nonumber \\
\end{equation}
\begin{equation}
\bar{\Pi}^{\rm RPA}_1(\bfq,\mathit{\omega})=\frac{-\mathrm{2}}{(\mathrm{2}
\mathit{\pi})^d}\int d\bfp\;
\frac{A^\ast(\bfp,\bfq)(\it{n}_{\rm 0}(\bfp+\bfq/\rm{2})-\it{n}_{\rm 0}
(\bfp-\bfq/\rm{2}))}
{\omega +i\gamma -E_c(\bfp+\bfq/\rm{2})+\it{E_c}(\bfp-\bfq/\rm{2})}.
\nonumber \\
\end{equation}
The dynamical dielectric function,
$\varepsilon(\bfq,\omega)$, is the same as defined in Eq. (7) within
the same RPA formulae (Eq. (8)). 
Note that resonance effects arising from $A(\bfp,\bfq)$
(i.e. considering the full two step process involving 
both conduction and valence bands rather than just the effective single step 
process (step 3 of Fig. 1(a)) within the conduction band) 
are nonperturbative and depend crucially 
on the exact value of the incident photon energy.
In the nonresonant theory, by contrast, the incident photon energy does not
enter into the calculation of the spectra, only the frequency shift
$\omega$ matters.
\section{Results and Discussions}
In Figs. 3 and 4 we show the energy 
dispersion and the dynamical structure factor respectively
of the nonresonant RRS
spectra in the RPA theory for 1D, 2D, and 3D semiconductor GaAs systems.
We emphasize that all earlier theoretical works on RRS spectroscopy, with 
the only exception of our earlier brief communication \cite{Dassarma99},
use the nonresonant approximation.
The sold lines in Fig. 4 are the RRS spectrum profiles 
in the long wavelength limit
(small momentum transfer, $|\bfq|$=0.1$k_F$), while the dashed lines
are the results of larger momentum transfer for comparison.
(The experimental situations correspond to the long wavelength limit,
with $|\bfq|\ll k_F$.)
Two elementary excitations are observed in the nonresonant spectra (Fig. 4)
at two separate peaks:
one is the single particle excitation at lower energy and the other one is
collective charge density
excitation at higher energy.
(Note that we use very small damping, $\gamma=10^{-3}E_F$, in
Fig. 4 in order to resolve the small SPE weights; larger $\gamma$ 
smears out the SPE continuum completely.)
We first mention that the RPA calculated energy dispersions 
of both modes (SPE and CDE) agree quantitatively
with the experimental RRS results \cite{6,11,13,14,hubbard,20}. 
However, the theoretically calculated nonresonant dynamical 
structure factor in Fig. 4 is entirely dominated by
the collective CDE mode, the SPE mode, while being present in the 
results, carries negligible and unobservable spectral weight. This
is entirely
inconsistent with the "two-peak" structure observed
in the experimental RRS spectra \cite{6} where the two peaks carry comparable
spectral weights. 
In the large momentum transfer results (which are outside the experimentally
accessible regime) shown
in Fig. 4 (dashed lines), one finds that 
SPE spectral weights are somewhat enhanced over the long wavelength results,
and correspondingly CDE weights decrease
for large momentum scattering 
due to the strong Landau
damping of plasmons (CDE) to single particle excitations which become
allowed at large wavevectors. The SPE spectral weight
is still much weaker (by three orders of magnitude) than the
CDE weight even at large wavevectors,
and in addition, the incoherent SPE continuum is 
severely broadened in this large
momentum scattering channel. Note that this situation (i.e.
negligible theoretical spectral weight at SPE) does not change 
\cite{7,wang01,16,21} even if
one goes beyond RPA and includes vertex corrections (e.g. Hubbard
approximation or time-dependent local density approximation) 
in the irreducible response function. Therefore, as long as resonance
effects are neglected (and thus one includes only
the step 3 of Fig. 1(a) ignoring the
interband resonance process), the calculated RRS spectra at experimentally
accessible wavevectors produce only observable 
CDE peaks in contrast to the experimental
two-peak situation which, in addition, always finds at resonance 
SPE spectral weight to be 
comparable to the CDE spectral weights 
\cite{6,rrs_exp96,Jusserand00,1,2,3,4,5,7,8,10,14,19}. 
The nonresonant theory is therefore in qualitative
disagreement with experiments as it fails to account the observed 
two-peak RRS spectra.

In Figs. 6-7 we show our results for the 
polarized RRS spectroscopy of the same 1D, 2D and 3D systems as in Fig. 4
within the resonant RPA theory (Eqs. (13)-(19))
in the long wavelength region ($|\bfq|$=0.1 $k_F$). 
RRS spectra for different resonance conditions, i.e. for different
values of $E_\omega$, are shown 
in Fig. 6 with a larger value of the impurity broadening
parameter ($\gamma=0.05E_F$, fifty times 
greater than the $\gamma$ used in Fig. 4) in order to compare with the 
experimental RRS profiles. The lower(higher) energy peak is associated 
with the SPE(CDE) of the electron systems.
The most important qualitative feature of the resonant 
theory results is the great enhancement of the SPE spectral 
weight compared with the nonresonant theory.
The three figures, Figs. 6(a), 6(b),
and 6(c) (corresponding to the results of 1D , 2D, and 3D systems respectively)
have qualitatively very similar behaviors: 
(i) the overall spectral weights decay very fast 
off resonance (i.e. for large $|E_\omega|$);
(ii) the peak positions of the SPE and CDE in Fig. 6
are the same as the nonresonant excitation energies in Fig. 4, i.e. 
resonance does not affect
the energy dispersion of the elementary electronic excitations;
(iii) the spectral weight
of SPE (lower energy peak) is essentially zero far away from resonance
($|E_\omega|>0.2$) where the CDE (higher energy peak) dominates similar to
the nonresonant spectra in Fig. 4 (except for the larger value of $\gamma$
used in Fig. 6); (iv) near resonance 
($|E_\omega|<0.2$) the SPE spectral weight is greatly enhanced ---
in fact, the SPE spectral weight becomes
comparable to or even larger than the CDE spectral weight, in sharp
contrast to the nonresonant theory (where the SPE weight is always extremely
small at long wavelength).
In Fig. 7 we plot our calculated RRS spectral weight ratio of
CDE/SPE as a function of the resonance condition, explicitly showing
the dramatic effect of resonance on the SPE spectral weight.
We emphasize that this spectacular
enhancement of SPE spectral weight in the full two step resonant scattering
process (over the simple one step nonresonant effective theory) 
is a nonperturbative
effect in our theory.
Our calculated spectra at resonance are in excellent qualitative
agreement with the corresponding experimental RRS spectra shown in
Ref.\cite{6,rrs_exp96,Jusserand00}, where the SPE spectral weight 
dies off rather quickly as the incident photon energy goes off-resonance.
From our results presented in Fig. 7, we also find that the
spectral weight ratio of the CDE to the SPE has very similar resonance 
behaviors
for systems of different dimensionalities, 
consistent with the experimental findings and indirectly
ensuring the validity of RPA theory in the RRS spectroscopy,
at least in the experimental parameter regimes. 

To understand the resonance condition dependence (on $E_\omega$) of Fig. 7,
we should explain the resonance effects not only on the SPE continuum,
but also on the CDE modes
around the resonance region. 
In some sense the extreme resonance condition,
$E_{\omega}=0$, may be thought of as providing an indirect mechanism for the
breakdown of the wavevector conservation for the scattering process
considered only within the conduction band in the prevailing
nonresonant theory where the virtual valence band effects are ignored
(i.e. step 3 in Fig. 1(a)) ---
thus our theory preserves the essence of the "massive" wavevector breakdown
mechanism proposed in Ref. \cite{2},
but in a very indirect sense because no impurity scattering is involved.
Instead, participation by the valence band introduces the effective
mechanism for wavevector conservation "breakdown" through
virtual interband process not included in the nonresonant theory.
In particular, the function $A(\bfp,\bfq)$ defined in Eq. (15)
provides the "wavevector conservation breaking" mechanism by mixing conduction
and valence band wavevectors non-trivially; if $A(\bfp,\bfq)$ is a constant,
there is no resonant enhancement of the SPE mode.
Equivalently, the dependence of $A(\bfp,\bfq)$ on two different wavevectors
is the effective wavevector conservation breakdown mechanism.
Mathematically we can start 
from the RPA dynamical structure factor
defined in Eq. (16), where the
CDE spectral weight is given by the numerator of the second term,
$\Pi_1^{\rm RPA}(\bfq,\mathit{\omega})\bar{\rm \Pi}_{\rm 1}^{\rm RPA}
(\bfq,\mathit{\omega})V_c(\bfq)$, at the CDE dispersion 
energy determined by the zero of the dielectric function 
($\varepsilon(\bfq,\omega)=\rm 0$).
Off-resonance, the function $A(\bfp,\bfq)$ is just a 
slowly varying function of momentum, $\bfp$, in the integral range 
$|\bfp\pm k_F|<\bfq/\rm{2}$ obtained by the occupancy factor,
$n_0(\bfp+\bfq/\rm{2})-\it{n}_{\rm 0}(\bfp-\bfq/\rm{2})$, in Eq. (17)-(19),
and therefore the RRS spectra show behavior (i.e. the CDE dominance
the SPE) similar to the standard RPA results 
(see Fig. 4) except for the overall decreasing weight factor, $E_\omega^{-2}$.
Near resonance ($E_\omega\sim 0$), however, the resonance function 
$A(\bfp,\bfq)$ in Eqs. (18) and (19)
can essentially cancel 
the contribution from the other integrand in the polarizabilities
(due to its sign change at $|\bfp|=k_F$),
so that the CDE spectral weight (coming 
essentially from the second term in Eq. (16))
cannot be as strongly enhanced by resonance 
as the SPE weight, which arises mostly from the
$\Pi_2^{\rm RPA}(\bfq,\omega)$ in Eq. (17).
Therefore the sign change of the resonant function, 
$A(\bfp,\bfq)$, is responsible for the relatively weaker enhancement of 
the CDE weight compared to the SPE weight near resonance.
We note that Eq. (16), defining the resonance spectral 
weight in our theory, has two terms, both of which are 
important in giving rise to a strong SPE spectral feature 
in the RRS spectra under resonance conditions.

Finally we give a simple explanation for the breakdown of Luttinger liquid
theory in the 1D RRS process near resonance. 
It is well known that 1D electron systems are
best understood as a Luttinger liquid, where
collective excitations are the only possible excitations
and no single particle excitations exist, for the
conduction band electrons. However, Luttinger liquid
behavior depends crucially on the charge conjugation 
symmetry, where the Hamiltonian remains the same after electrons and holes 
are exchanged about the Fermi surface.
When the valence band is intrinsically involved near resonance in
the RRS process, such electron-hole 
conjugation symmetry is totally broken, because the filled valence 
band is effectively "overlapped" with the conduction band at 
Fermi surface. In other words, an electron below the
conduction band Fermi surface now effectively 
has a new channel, not restricted by
the small 1D phase space, to be excited
above the conduction Fermi surface through 
the two step resonant interband transition, 
through the valence band virtual transition.
An estimated resonance condition for this apparent 
breakdown of LL behavior in 1D 
RRS spectroscopy can therefore be obtained by $|E_\omega|<qv_F/E_F=2(q/k_F)$,
which is 0.2 for $q=0.1 k_F$ and is consistent with our numerical result
shown in Fig. 7. We therefore physically 
explain the failure of the theoretical attempt
of using LL theory to study the 1D RRS experiments near resonance
\cite{Wang00}. 
The qualitative similarity of the experimental RRS results for
one, two and three dimensional systems confirms our theory, which is 
based on the conventional Fermi liquid model.
We note, however, that the Luttinger liquid description 
of 1D systems has recently been theoretically modified 
\cite{Wang00} in an attempt to understand the observed 
RRS spectra, but the applicable theory is quite subtle 
and beyond the scope of this paper.
\section{Summary}
In summary, it may be important to emphasize 
that the striking phenomenological
similarity in the experimentally observed RRS spectra in one, two, and three
dimensional systems is a strong indication that generic 
interband resonance physics
as studied here 
(within a \textit{resonant} RPA scheme) is playing
a fundamental role in producing the low energy "SPE" feature in the polarized
RRS spectra, which cannot be explained by the standard 
(nonresonant) theory or any other 
non-generic (system-dependent) theories. Our theory can also be
applied to the depolarized RRS experiments, where both single particle
and spin density excitations are important, but the exchange energy should
be included properly \cite{7,21} to separate these two excitations which
are degenerate in the regular RPA calculation.
Once exchange correlation effects are invoked to distinguish 
the SPE and the SDE (with the SDE lying below the SPE by the 
exchange energy), our resonant theory can account for the 
observed two-peak structure in the resonant depolarized 
RRS experiments in a way very similar to the theory developed 
herein for the SPE and CDE in the polarized RRS experiments.
To summarize our results, we have developed a 
theory for resonant Raman scattering spectroscopy 
in one, two and three dimensional
semiconductor structures by considering the full two step 
resonance process involved in the scattering of external photons. We find that 
at resonance the RRS spectra have considerable weight at the SPE energy with 
the SPE weight decreasing off resonance. There is no qualitative difference
in the RRS spectra between the systems of different dimensions.
Our results are in qualitative
agreement with experimental findings and provide a generic theoretical 
explanation for an ubiquitous puzzle which dates back more than twenty
five years. As a concluding note we point out that it may be somewhat
misleading to call the additional feature in the RRS spectra an "anomalous"
SPE mode as has routinely been done in the literature --- 
a pure SPE mode arises from the imaginary part of the 
irreducible polarizability function, as given within 
RPA by Eq. (8), whereas the anomalous additional RRS 
feature arises primarily from the presence of the 
$\Pi^{\rm RPA}_2$ term (Eq. (17)) in
our resonant RPA theory (Eqs. (16)-(19)) which is (related to, but)
quite different from the irreducible polarizability, $\Pi^{\rm RPA}_0$
(Eq. (8)) by virtue of the nontrivial nature of the resonance function
$A(\bfp,\bfq)$. Finally, we mention that a very recent experimental report
has appeared in the literature \cite{Jusserand00} specifically verifying 
the essential features of our theory \cite{Dassarma99}. However a complete
quantitative understanding of experimental results may very well require
inclusion of additional effects (e.g. excitonic corrections, many body 
effects) beyond the scope of our work.

We thank A. J. Millis for critical discussion. The work is supported
by the US-ARO, the US-ONR and DARPA.

\begin{figure}
 \vbox to 7.5cm {\vss\hbox to 5.5cm
 {\hss\
   {\includegraphics{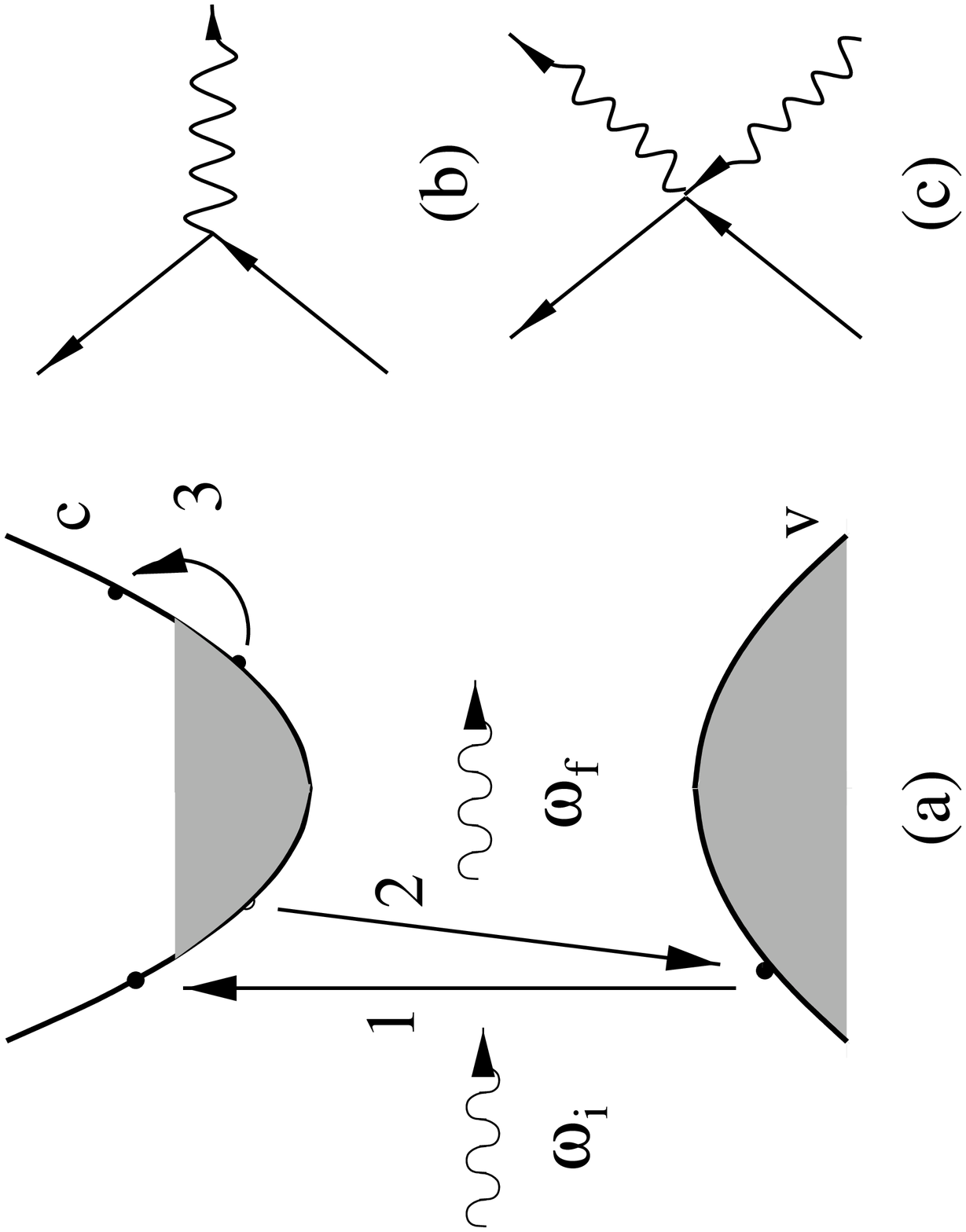}
   }
  \hss}
 }
\caption{(a) Schematic representation of RRS in the 
direct gap two band model of electron doped GaAs nanostructure. $\omega_i$ and 
$\omega_f$ are the initial and final frequencies of the external photons. 
Steps 1, 2 and 3 are described in the text (RRS involves steps 1 and 2 only).
(b) and (c) are the
Feynman diagrams of the electron-photon scattering process described by 
$\bfp\cdot\bfA$ and $\bfA\cdot\bfA$ terms respectively in the interacting 
Hamiltonian, $H_I$. Solid and wavy lines represent the electron and 
photon Green's functions respectively.
}
\label{fig:1}
\end{figure}
\begin{figure}
 \vbox to 6.8cm {\vss\hbox to 5.5cm
 {\hss\
   {\includegraphics{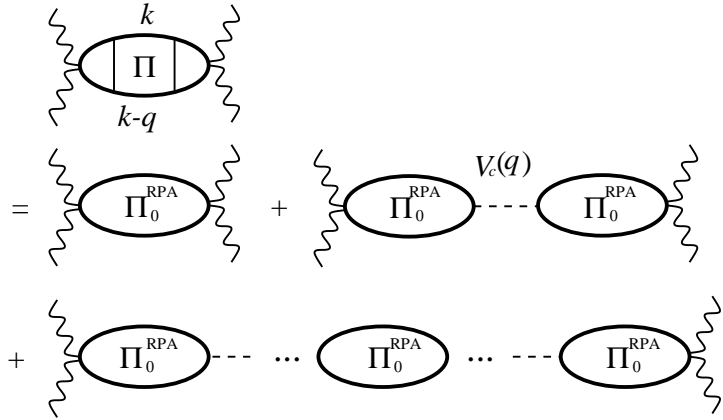}
   }
  \hss}
 }
\caption{ 
Diagrammatic representation of the conduction band irreducible response
function, $\Pi^{\rm RPA}_0(\bfq,\omega)$ and reducible response function, 
$\Pi(\bfq,\omega)$, in the standard random phase approximation.
$V_c(\bfq)$ is the
Coulomb interaction.
}
\label{fig:2}
\end{figure}
\begin{figure}
 \vbox to 6.5cm {\vss\hbox to 5.5cm
 {\hss\
   {\includegraphics{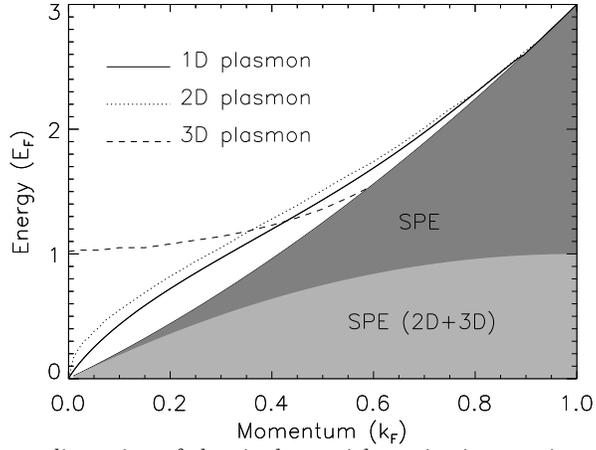}
   }
  \hss}
 }
\caption{Typical 
momentum-energy dispersion of 
the single particle excitation continuum (shaded region)
and the collective charge density excitations (plasmons) of one, two,
and three dimensional electron systems (calculated within RPA).
}
\label{fig:3}
\end{figure}
\begin{figure}
 \vbox to 6.1cm {\vss\hbox to 5.5cm
 {\hss\
   {\includegraphics{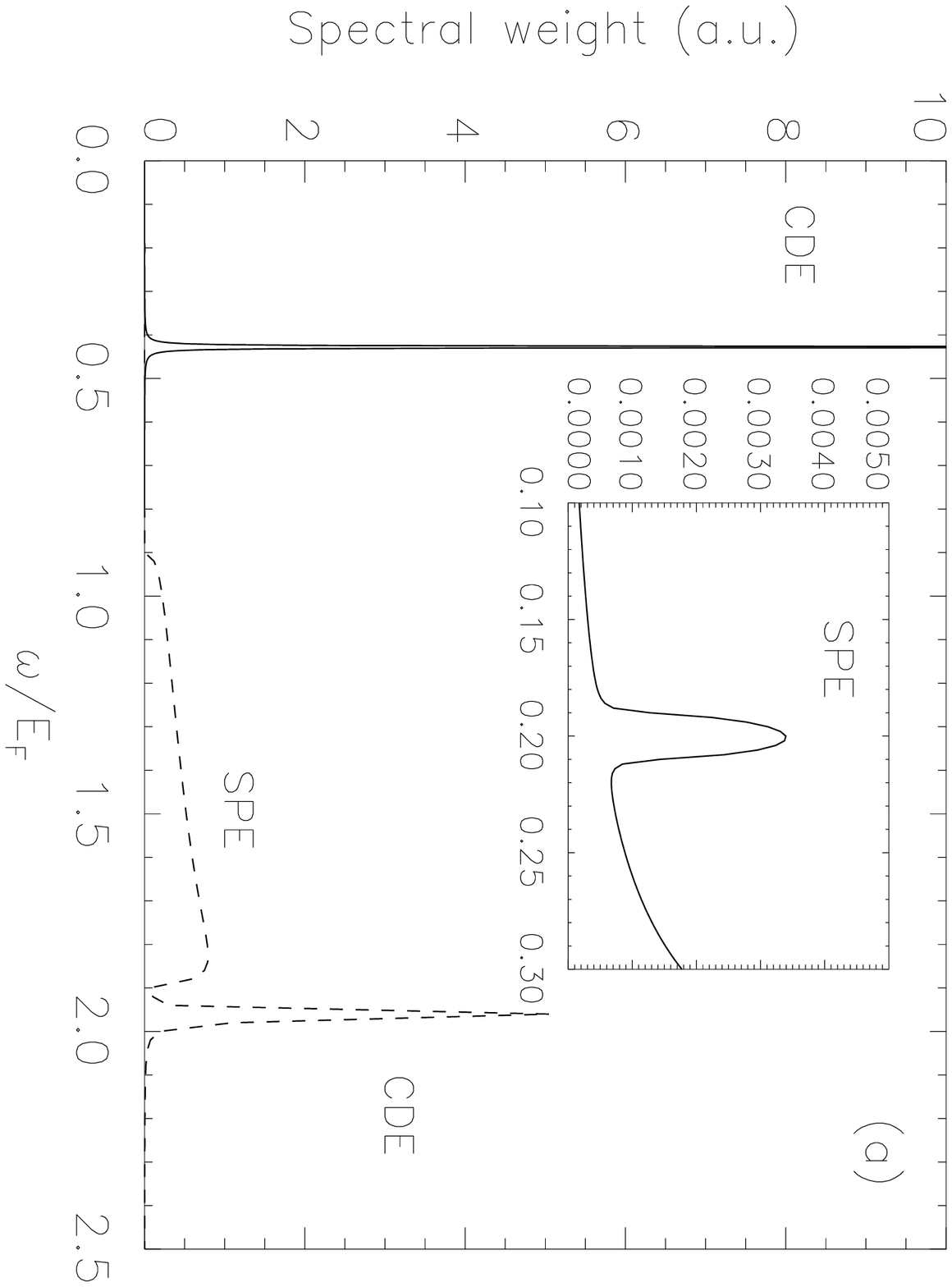}
   }
  \hss}
 }
 \vbox to 6.1cm {\vss\hbox to 5.5cm
 {\hss\
   {\includegraphics{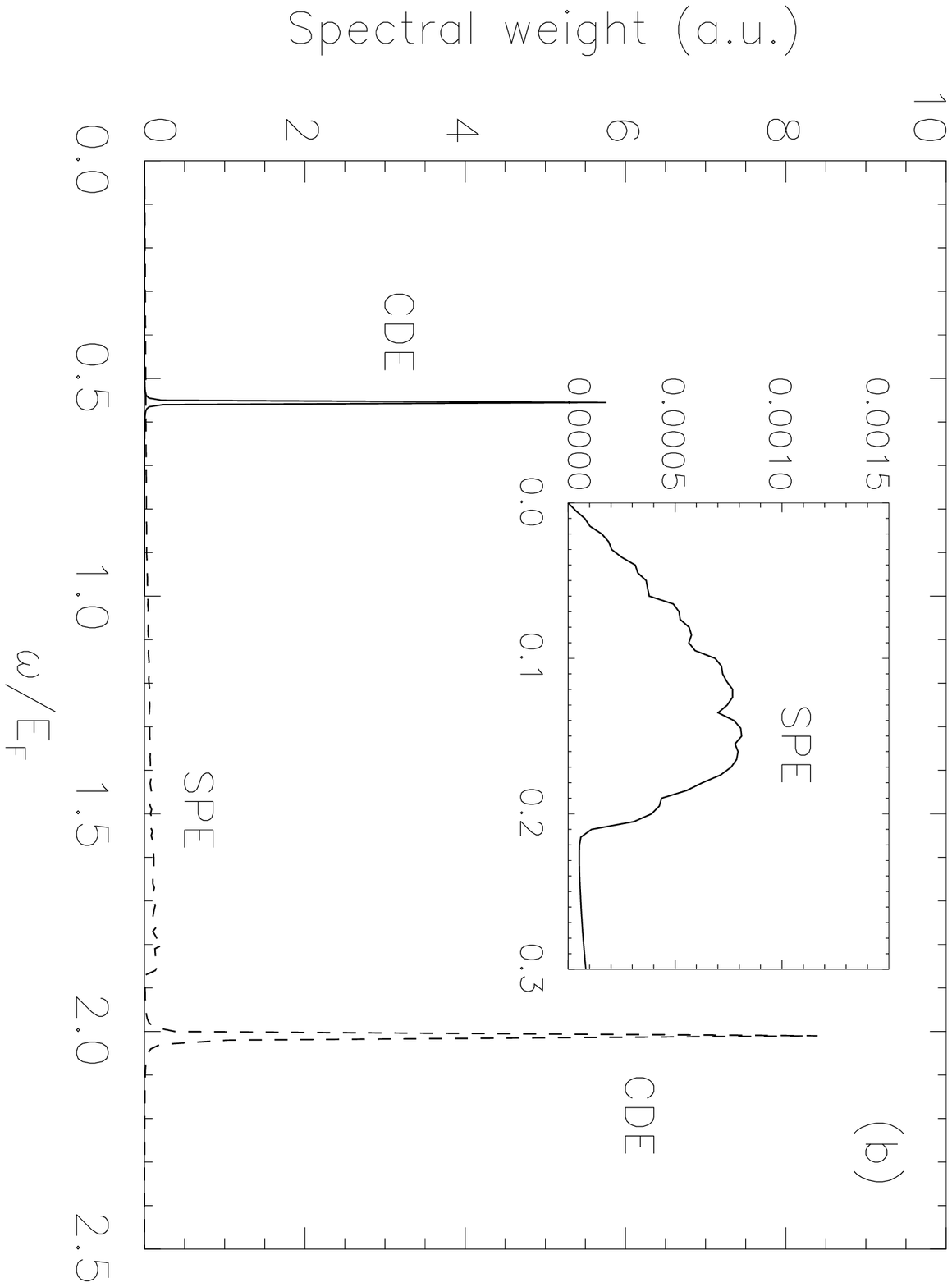}
   }
  \hss}
 }
 \vbox to 6.7cm {\vss\hbox to 5.5cm
 {\hss\
   {\includegraphics{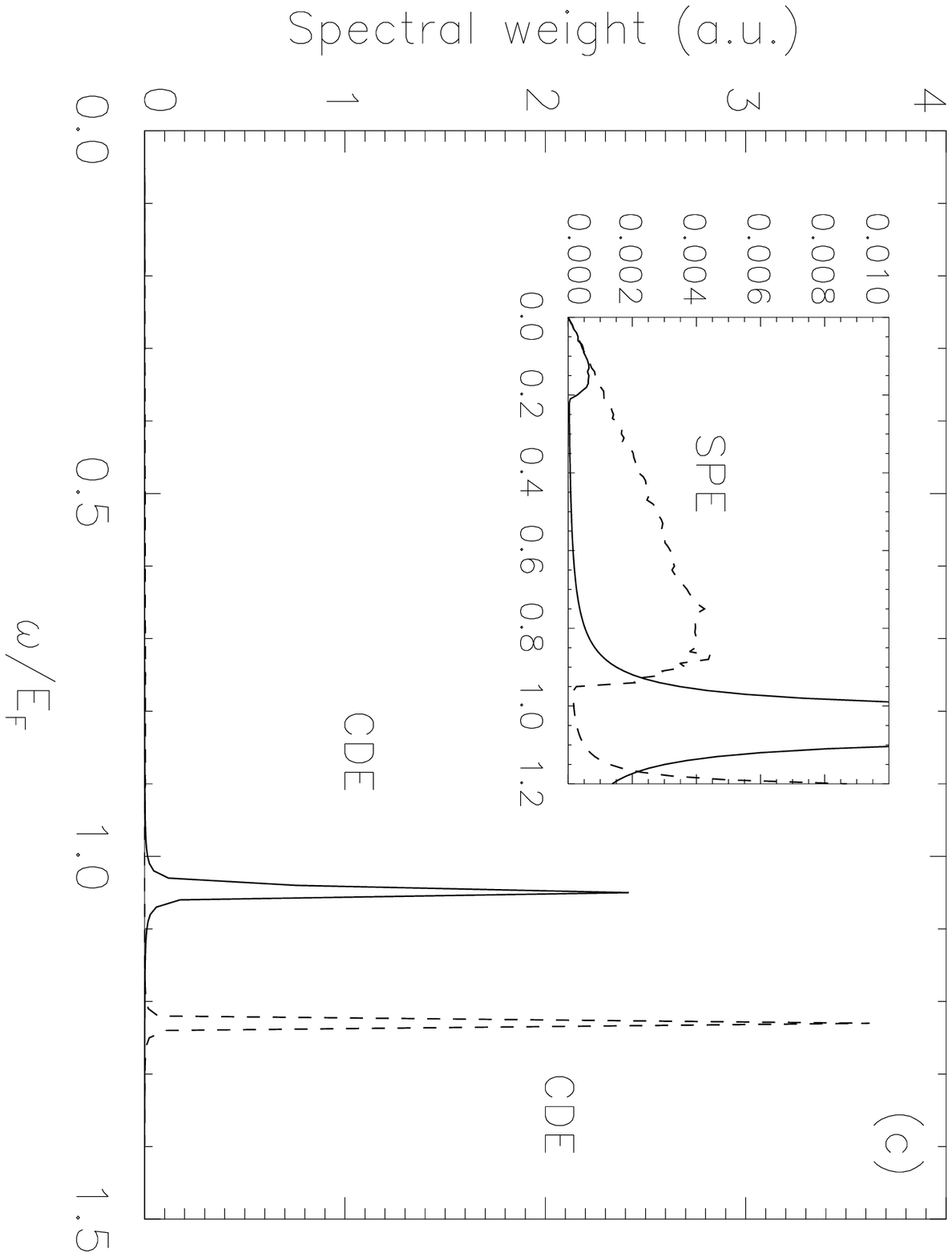}
   }
  \hss}
 }
\caption{
Dynamical structure factor obtained by the standard 
(nonresonant) RPA calculation at zero temperature
for (a) one, (b) two, and (c) three dimensional
electron systems ignoring valence band effects. Solid lines
are the calculated spectra in the 
long wavelength limit (small momentum transfer, 
$|\bfq|=\rm{0.1}\it k_F$)
and dashed lines are for the large momentum transfer
($|\bfq|=\rm{0.7}\it k_F$ in (a) and (b),
and $|\bfq|=\rm{0.4}\it k_F$ in (c)) calculations. The electron densities
used in the calculation is $6.5\times 10^5$ cm$^{-1}$, 
$3.2\times 10^{11}$ cm$^{-2}$, and $1.8\times 10^{17}$ cm$^{-3}$ for 
one, two, and three dimensional systems respectively. Finite impurity
scattering ($\gamma=10^{-3}E_F$) has been included to broaden the peaks.
The ripple in (b) and (c) is of numerical origin.
}
\label{fig:4}
\end{figure}
\begin{figure}
 \vbox to 8cm {\vss\hbox to 5.5cm
 {\hss\
   {\includegraphics{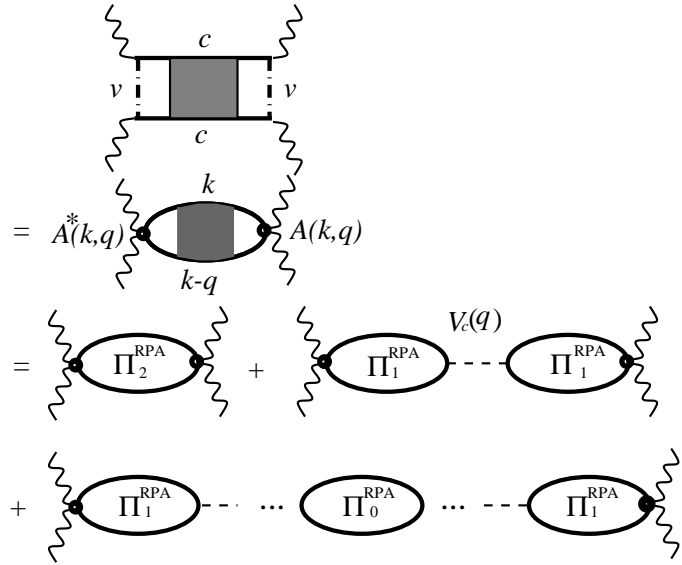}
   }
  \hss}
 }
\caption{
Diagrammatic representation of the resonant Raman scattering response 
function including the valence band electrons in the RPA calculation.
Different kinds of irreducible response functions are defined 
and explained in Eqs. (17)-(19) and the matrix element, $A(\bfk,\bfq)$,
is defined in Eq. (15). 
}
\label{fig:5}
\end{figure}
\begin{figure}
 \vbox to 8cm {\vss\hbox to 5.5cm
 {\hss\
   {\includegraphics{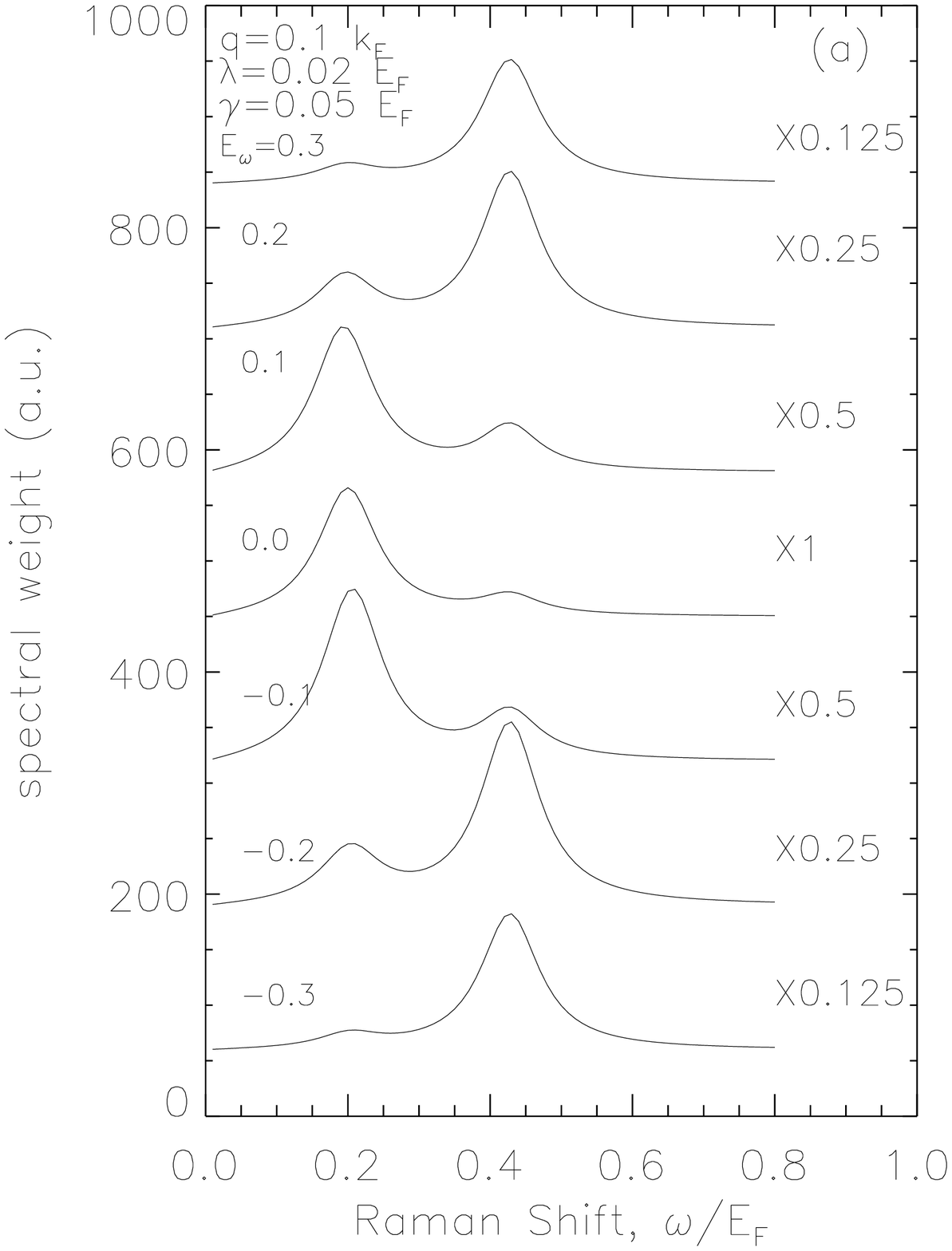}
   }
  \hss}
 }
 \vbox to 10cm {\vss\hbox to 5.5cm
 {\hss\
   {\includegraphics{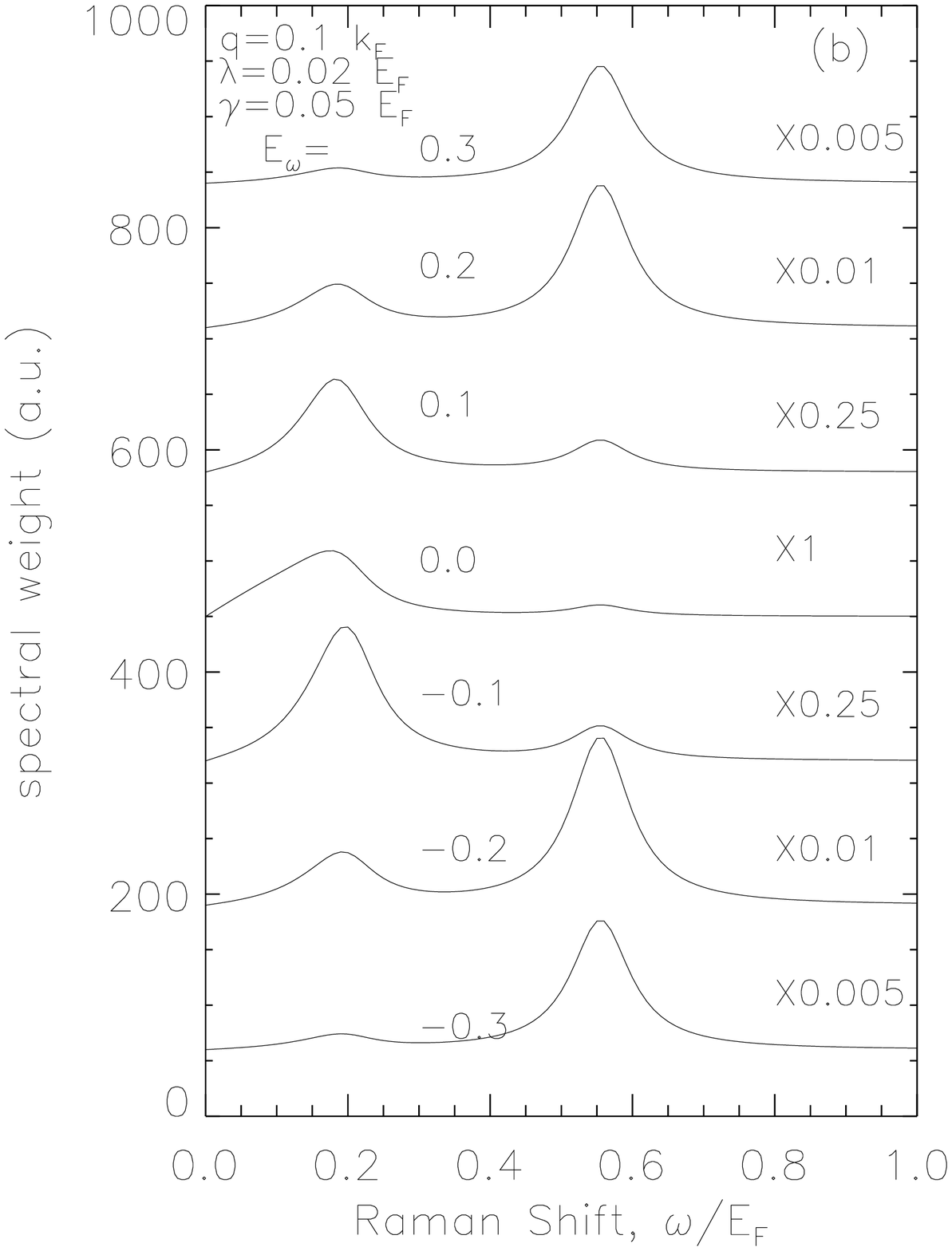}
   }
  \hss}
 }
 \vbox to 8cm {\vss\hbox to 5.5cm
 {\hss\
   {\includegraphics{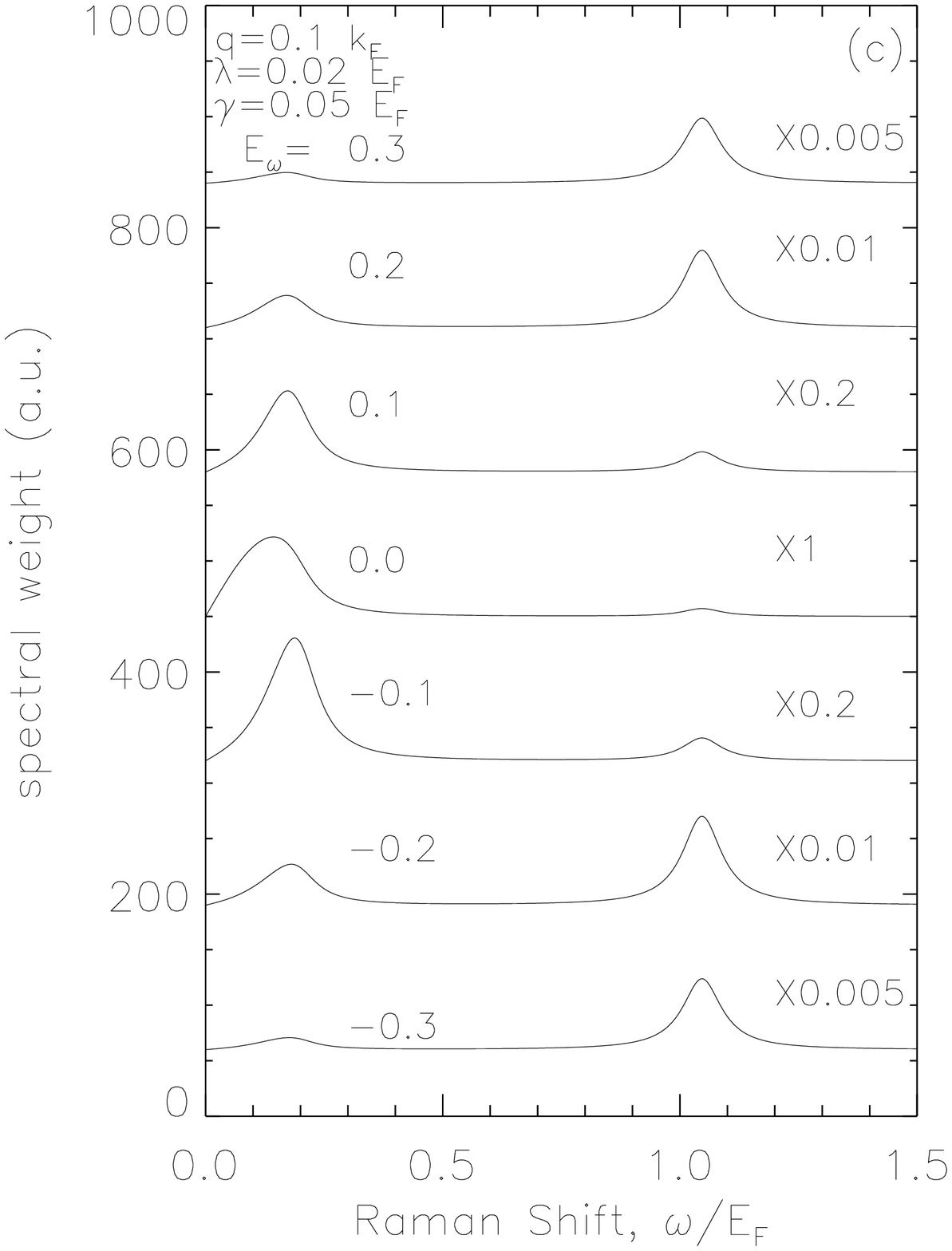}
   }
  \hss}
 }
\caption{
Dynamical structure factor in the resonant RPA calculation
for (a) one, (b) two, and (c) three dimensional
electron systems incorporating the valence band electrons. 
We choose the resonance broadening factor, $\lambda$, to be $0.02E_F$,
and finite impurity scattering factor, $\gamma$, to be 0.05 $E_F$
in order to have agreement with
the experimental data (the impurity broadening is still rather
small since $\gamma/E_F=1/20$). Other system parameters are the same
as used in Fig. 4.
}
\label{fig:6}
\end{figure}
\begin{figure}
 \vbox to 6.7cm {\vss\hbox to 5.5cm
 {\hss\
   {\includegraphics{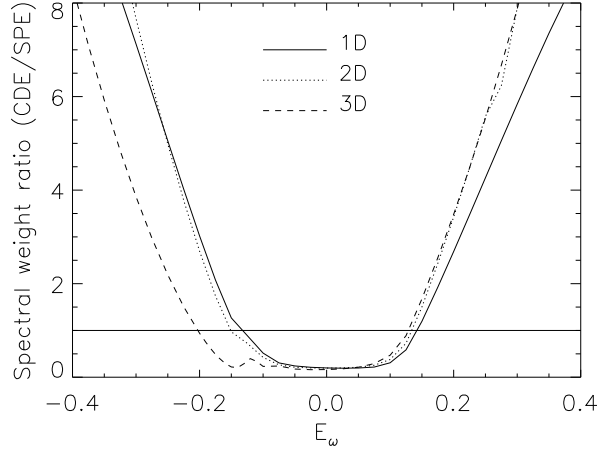}
   }
  \hss}
 }
\caption{
Ratio of the resonant Raman scattering spectral weight (CDE to SPE)
as a function of the resonance energy,
$E_{\omega}$, in one, two, and three dimensional
systems. Off-resonance, 
$|E_\omega|\geq 0.2$, CDE always dominates SPE in the spectra, but  
near resonance,
$|E_\omega|<0.2$, the SPE weight could 
even be stronger than the CDE weight. All system parameters 
are the same as used in Fig. 6. 
}
\label{fig:7}
\end{figure}
\end{document}